\documentclass[12pt,amsmath,amssymb]{revtex4}
\usepackage{amsfonts}
\usepackage{amsmath}
\usepackage{amssymb}
\usepackage{array}
\usepackage{bm}
\usepackage{braket}
\usepackage{breqn}
\usepackage{color}
\usepackage{dcolumn}
\usepackage{epsfig}
\usepackage{epsf}
\usepackage{epstopdf}
\usepackage{float}
\usepackage{graphicx}
\usepackage{graphics}
\usepackage{harpoon}
\usepackage{indentfirst}
\usepackage{mathrsfs}
\usepackage{multirow}
\usepackage{pifont}
\usepackage{xcolor}

\newcommand {\sla}[1]{ #1 \!\!\!/}

\newcommand{\RM}[1]{\textrm{\uppercase\expandafter{\romannumeral#1}}}

\newcommand{\beq}{\begin{eqnarray}}
\newcommand{\eeq}{\end{eqnarray}}

\begin{document}

\title{Two-Photon-Exchange effect in $ep\rightarrow en\pi^+$ at small $-t$ with the hadronic model and dispersion relation approach}
\author{
Qian-Qian Guo, Hai-Qing Zhou \protect\footnotemark[1] \protect\footnotetext[1]{E-mail: zhouhq@seu.edu.cn} \\
School of Physics,
Southeast University, NanJing 211189, China}
\date{\today}

\begin{abstract}
In this work, the two-photon-exchange (TPE) effect in $ep\rightarrow en\pi^+$ at small $-t$ is discussed. In the previous work, the TPE contribution with one $\pi$ intermediate state is estimated numerically within a hadronic model under the pion-dominance approximation. Here we extend the discussion to include one $\rho$ intermediate state. The TPE contribution can be described by one scalar function in the limit $m_e\rightarrow 0$, the dispersion relation (DR) satisfied by this scalar function is analysed. The analytic expressions for the imaginary parts of the TPE contributions from one $\pi$ or one $\rho$ intermediate state are given within the hadronic model. Combining these analytic expressions and the DR, the corresponding real parts of the TPE contributions can be estimated easily at any available region. This can help the further experimental analysis to include the TPE contributions in a convenient way. The numeric results show that the TPE correction with one $\rho$ intermediate state is much smaller than that with one $\pi$ intermediate state in the current energy region. These results suggest that the TPE contribution with an elastic state is the main TPE contribution in $ep\rightarrow en\pi^+$ at small $-t$.
\end{abstract}

\maketitle


\section{Introduction}

The two-photon-exchange (TPE) effect palys an important role to extract the the electromagnetic (EM) form factors (FFs) of the proton from the unpolarized $ep$ scattering and has been widely studied by many theoretical methods such as the hadronic model \cite{hadronic model}, GPD method \cite{GPD method}, pQCD calculation \cite{pQCD method}, dispersion relation (DR) approach \cite{dispersion relation-1}\cite{dispersion relation-2}, soft collinear effective theory (SCEF) method \cite{SCEF}, chiral perturbative theory (ChpT) \cite{TPE-ep-ChpT-2020-2021} and phenomenological parametrization \cite{phenomenological parametrizations}. Recently many experimental measurements are developed to test these theoretical estimations and to deep our understanding on the TPE contribution\cite{VEPP3-2012,VEPP3-2014,Moteabbed13,Kohl14}.

Comparing with the proton case, the discussions on how to extract the EM FF of $\pi$ precisely are relatively fewer. Experimentally, the EM form factor of $\pi$ is usually extracted  via the process $ep\rightarrow en\pi^+$ \cite{electro-pion-production-Cornell,electro-pion-production-DESY,electro-pion-production-JLab-1,
electro-pion-production-JLab-2,electro-pion-production-JLab-3}. Theoretically, such extraction of pion's FF is much more complex than that of the proton's FFs via the elastic $ep$ scattering. The corresponding theoretical analysis on the experimental data sets should be done more carefully. Up to now, the discussions on the TPE effect in $ep\rightarrow en\pi^+$ are limited \cite{Afanasev2013,zhouhq2020-TPE-pi-intermediate}. In the previous work \cite{zhouhq2020-TPE-pi-intermediate}, the TPE contributions with an elastic intermediate state are discussed, in this work we extent the discussion to include one $\rho$ meson intermediate state. Furthermore, the DR for the TPE contributions and the analytic expressions for the imaginary  parts are both given.

We organize the paper as follows. In Sec.  II we describe the basic frame of our discussion under the pion-dominance approximation, in Sec. III we show some analytic properties of the TPE contributions and the DR relation they satisfied. in Sec. IV we present some numerical results for the TPE corrections and give our conclusion.

\section{Basic Frame for The TPE Contributions in $ep\rightarrow en\pi^+$}

\begin{figure}[htbp]
\centering
\includegraphics[height=5cm]{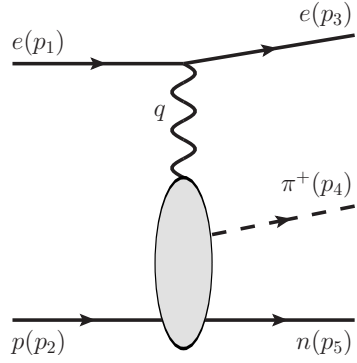}
\caption{$ep\rightarrow en\pi^+$ with one-photon exchange.}
\label{Fig:ep-enpi-OPE-general}
\end{figure}

Under the one-photon exchange (OPE) approximation, the process $ep\rightarrow en\pi^+$  can be described by Fig. \ref{Fig:ep-enpi-OPE-general} where we label the momenta of initial electron, initial proton, final electron, final pion and final neutron as $p_{1,2,3,4,5}$. For simplicity we define the following five independent Lorentz invariant variables $s \equiv (p_1+p_2)^2$, $Q^2\equiv-q^2\equiv -(p_1-p_3)^2$, $W\equiv\sqrt{(p_4 + p_5)^2}$,   $t\equiv (p_2-p_5)^2$ and  $\nu=(p_1+p_3) \cdot(2p_4+p_3-p_1)$.

\begin{figure}[bhtp]
\centering
\includegraphics[height=4cm]{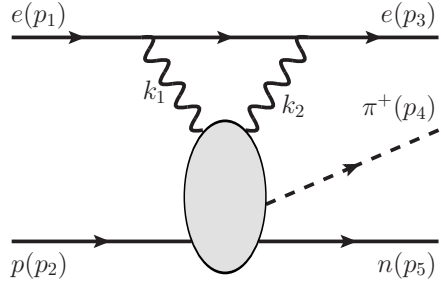}
\caption{$ep\rightarrow en\pi^+$ with two-photon exchange.}
\label{Fig:ep-enpi-TPE-general}
\end{figure}

When go to discuss the TPE effect, the contribution from the corresponding TPE diagram showed in Fig. \ref{Fig:ep-enpi-TPE-general} should be considered.

\begin{figure}[bhtp]
\centering
\includegraphics[height=4cm]{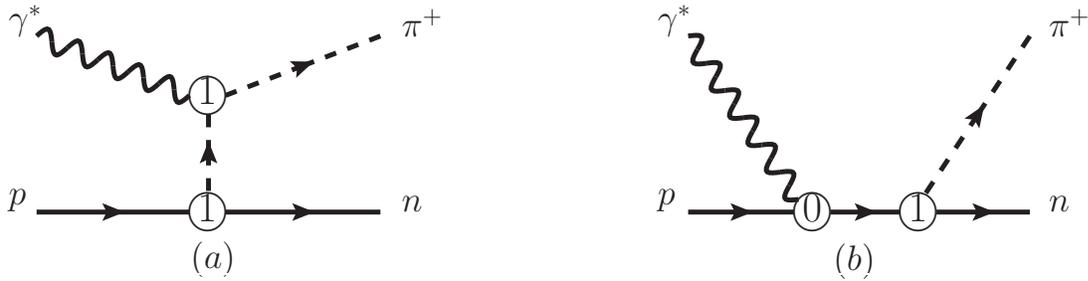}
\caption{Diagrams for $\gamma^*p\rightarrow n\pi^+$  in the leading order of ChpT with ($a$) the pion exchange diagram and ($b$) the elastic $s$-channel diagram.}
\label{Fig:gammap-npi-LO}
\end{figure}

Physically, the dynamics of the sub-processes $\gamma^*p\rightarrow n\pi^+$ and $\gamma^*\gamma^*p\rightarrow n\pi^+$ are very complex.
At the small energy scale, one can expect that the chiral perturbative theory (ChpT) works well for these two sub-processes. For example, in the leading order of ChpT the Feynman diagrams for $\gamma^*p\rightarrow n\pi^+$ can be described  by Fig.\ref{Fig:gammap-npi-LO} where the notations $\textcircled{0}$ and $\textcircled{1}$ refer to the vertexes with corresponding orders.

\begin{figure}[htbp]
\centering
\includegraphics[height=4cm]{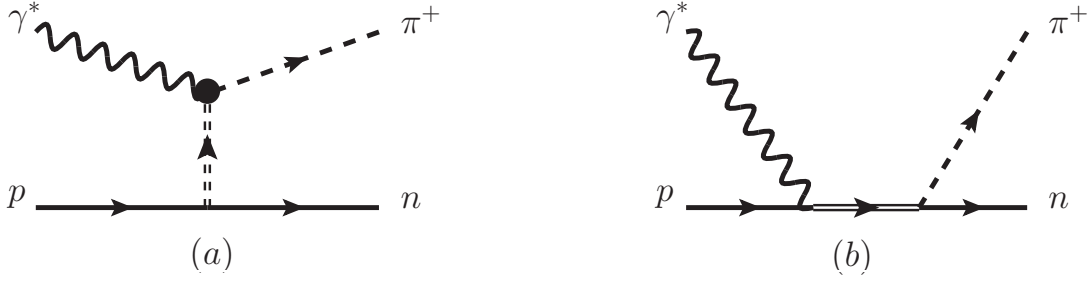}
\caption{Examples of diagrams for $\gamma^*p\rightarrow n\pi^+$  beyond ChpT with ($a$) $\rho$ meson exchange and ($b$)  $N^*$ contribution.}
\label{Fig:gammap-npi-rho-Nstar}
\end{figure}

When the energy scales $-t,Q^2$ and $W$ increase, one can expect that ChpT is not valid anymore and the contributions beyond ChpT such as the diagrams with $\rho$ meson exchange and with $N^*$ intermediate states showed as Fig. \ref{Fig:gammap-npi-rho-Nstar} should be considered.

When $-t$ is kept as small, $W$ is a little far away from the masses of the narrow resonances and only $Q^2$ increases, the contribution from one pion-exchange is still dominant among all these contributions although ChpT is not valid. The reasons are due to two facts: (1) the mass of pion is close to zero which results in a strong enhancement from the pion propagator, (2) the couplings of $\gamma^*pp$,$\gamma^*pN^*$ decrease much fast than the coupling $\gamma^*\pi\pi$ when $Q^2$ increases. These properties means the pion-dominance is a good approximation when $t\rightarrow 0$ and $W$ is a little far away from the narrow resonances. This also greatly simplifies the dynamics of the process  $\gamma^*\gamma^* p\rightarrow n\pi^+$ in this region. In this work, we limit our discussion on the TPE contributions under this approximation. In the practical calculation, one can combine the contributions beyond the pion-dominance under the OPE approximation and the TPE contributions together since their contributions are independent.

\begin{figure}[htbp]
\centering
\includegraphics[height=4cm]{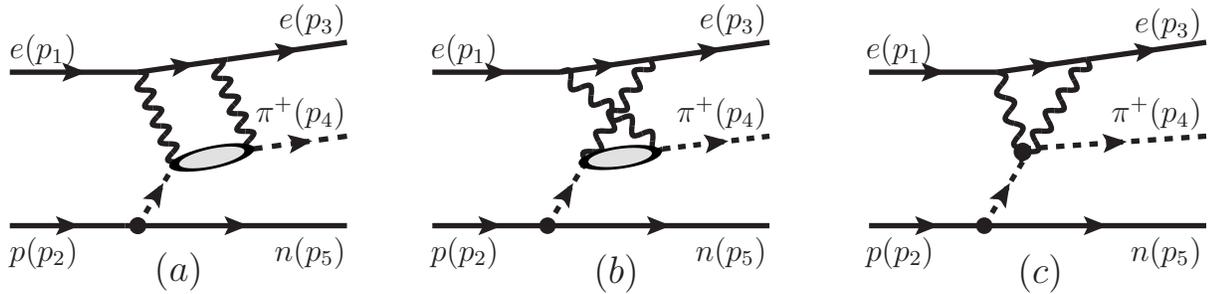}
\caption{Diagrams for $ep\rightarrow en\pi^+$  with two-photon exchange under the pion-dominance approximation: ($a$) is the box diagram, ($b$) is the crossed-box diagram and ($c$) is the contact diagram.}
\label{Fig:ep-enpi-TPE-abc-general}
\end{figure}

\begin{figure}[htbp]
\centering
\includegraphics[height=3.5cm]{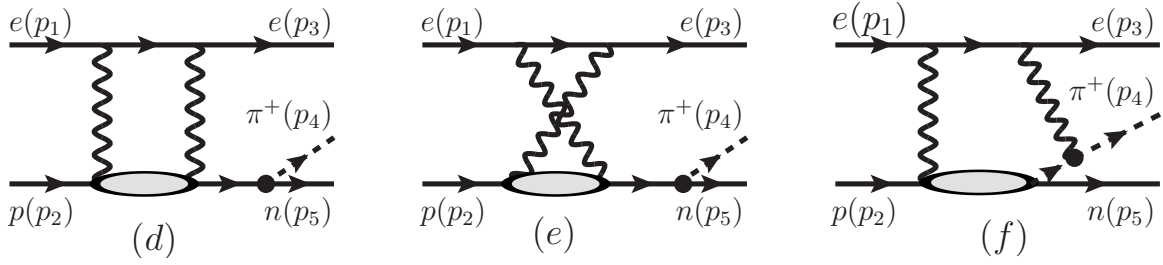}
\caption{Diagrams for $ep\rightarrow en\pi^+$  with two-photon exchange beyond the pion-dominance approximation.}
\label{Fig:ep-enpi-TPE-def-general}
\end{figure}

Under the pion-dominance approximation, the corresponding TPE contributions can be described as Fig. \ref{Fig:ep-enpi-TPE-abc-general}($a,b,c$) where the contributions from Fig. \ref{Fig:ep-enpi-TPE-def-general}($d,e,f$) are neglected since they are much smaller.

\begin{figure}[htbp]
\centering
\includegraphics[height=3.8cm]{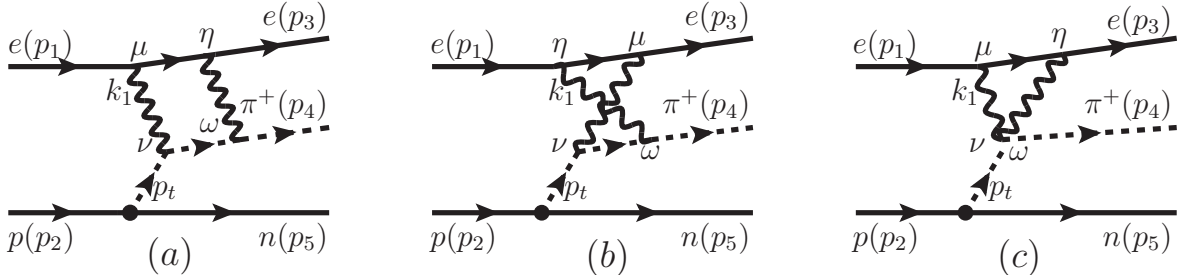}
\caption{Diagrams for $ep\rightarrow en\pi^+$  with two-photon exchange and one $\pi$ meson intermediate state: ($a$) is the box diagram ($b$) is the crossed diagram and ($c$) is the contact diagram.}
\label{Fig:ep-enpi-TPE-pi-intemediate-state}
\end{figure}

\begin{figure}[htbp]
\centering
\includegraphics[height=4.2cm]{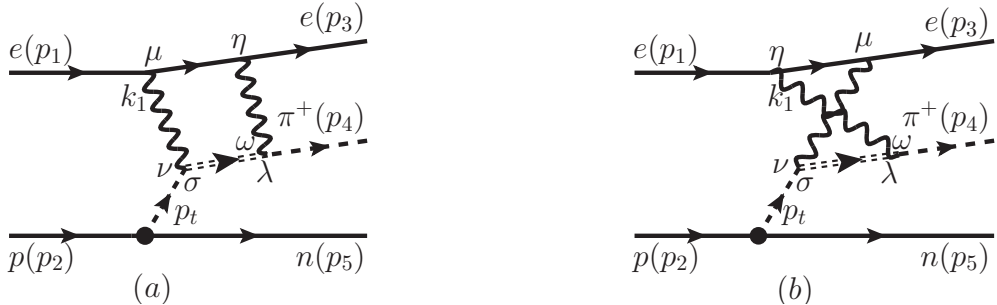}
\caption{Diagrams for $ep\rightarrow en\pi^+$  with two-photon exchange and one $\rho$ meson intermediate state: ($a$) is the box diagram and ($b$) is the crossed-box diagram.}
\label{Fig:ep-enpi-TPE-rho-intemediate-state}
\end{figure}

In the previous work \cite{zhouhq2020-TPE-pi-intermediate},  the TPE contributions from an elastic state $\pi$ showed in Fig. \ref{Fig:ep-enpi-TPE-pi-intemediate-state} are discussed in the region $Q^2\subseteq [1,2.45]$GeV$^2$. Naively, at higher $Q^2$, the similar contributions with one $\rho$ meson intermediate state showed as Fig.  \ref{Fig:ep-enpi-TPE-rho-intemediate-state} should be considered.

Taking Feynamn gauge, one has
\begin{eqnarray}
\mathcal{M}_{1\gamma}^{(a)} &=& - i  \bar{u}_e(p_3)(-ie\gamma^{\mu}) u_e(p_1)\  \bar{u}_n(p_5)\Gamma_{5} u_p(p_2)\Gamma^{\nu}(p_4,p_t) S_{\pi}(p_t)D_{\mu \nu}(p_1-p_3), \nonumber \\
\mathcal{M}_{2\gamma,\rho}^{(a)} &=& - i \int \frac{d^4 k_1}{(2\pi)^4}\bar{u}_e(p_3)(-ie\gamma^{\eta})S_F(p_1-k_1)(-ie\gamma^{\mu}) u_e(p_1)\ \bar{u}_n(p_5)\Gamma_{5} u_p(p_2)] \nonumber \\
&&~~~~~~ \times \Gamma^{\omega \lambda}(p_1-k_1-p_3,p_t+k_1)S^{V}_{\lambda \sigma}(p_t+k_1)\Gamma^{\nu \sigma}(k_1,-p_t-k_1) ] \nonumber \\
&&~~~~~~ \times S_{\pi}(p_t)D_{\eta \omega}(p_1-k_1-p_3)D_{\mu \nu}(k_1), \nonumber \\
\mathcal{M}_{2\gamma,\rho}^{(b)} &=& - i \int \frac{d^4 k_1}{(2\pi)^4}\bar{u}_e(p_3)(-ie\gamma^{\mu})S_F(p_1-k_1)(-ie\gamma^{\eta}) u_e(p_1)\ \bar{u}_n(p_5)\Gamma_{5} u_p(p_2) ] \nonumber \\
&&~~~~~~ \times \Gamma^{\omega \lambda}(k_1,p_4-k_1)S^{V}_{\lambda \sigma}(p_4-k_1)\Gamma^{\nu \sigma}(p_1-k_1-p_3,k_1-p_4) ]  \nonumber \\
&&~~~~~~ \times S_{\pi}(p_t)D_{\mu \nu}(p_1-k_1-p_3)D_{\eta \omega}(k_1),
\label{Eq:amplitudes-OPE-TPE}
\end{eqnarray}
with $p_t=p_3+p_4-p_1$ and
\begin{eqnarray}
S_F(k) &=& \frac{i(\sla{k}+m_e)}{k^2-m_e^2+i\epsilon}, \nonumber \\
S_{\pi}(k) &=& \frac{i}{k^2-m_\pi^2+i\epsilon}, \nonumber \\
S^{V}_{\lambda \sigma}(k) &=&  \frac{- i(g_{\lambda\sigma}- \frac{k_{\lambda} k_{\sigma}}{k^2})}{k^2-m_{\rho}^2+i\epsilon}, \nonumber \\
D_{\mu\nu}(k) &=& \frac{-i}{k^2+i\epsilon}g_{\mu\nu},
\end{eqnarray}
and
\begin{eqnarray}
\Gamma^{\mu}(p_f,p_i) &=& ie[(1+f(k^2)k^2)(p_f+p_i)^{\mu}-f(k^2)(p_f^2-p_i^2)k^{\mu} ], \nonumber\\
\Gamma^{\mu\nu}(q_\gamma,q_V) &=& ie g_{\rho\pi\gamma}/m_{\rho}F_{\gamma\pi\rho}(q_{\gamma}^2)\epsilon_{\mu\nu\lambda\omega}q_{\gamma}^{\lambda}q_{V}^{\omega},
\end{eqnarray}
where $e=-|e|$, $k\equiv p_f-p_i$,$q_{\gamma,V}$ are the incoming momenta of photon and $\rho$ meson, $f(k^2)$ describes the EM form factor of pion $F_\pi(k^2)$ and has the relation\cite{zhouhq2011-pion-photon-interaction}
\begin{eqnarray}
F_{\gamma\pi\pi}(k^2)&=&1+k^2f(k^2).
\label{Eq:relation}
\end{eqnarray}
Here $F_{\gamma\pi\rho}(q_\gamma^2)$ is the transition FF of $\gamma^*\pi\rightarrow\rho$ \cite{g-rho-pi-gamma}.
The similar expressions $\mathcal{M}_{2\gamma,\pi}^{(a,b,c)}$  corresponding  to Fig. \ref{Fig:ep-enpi-TPE-pi-intemediate-state}($a,b,c$)  can be found in Ref. \cite{zhouhq2020-TPE-pi-intermediate}.  In the practical calculation, one can find that the relative TPE corrections are not dependent on the form of $\Gamma_5$ (iso-scalar form or iso-vector form), so we do not present its form here.

Generally, the amplitudes given in Eq. (\ref{Eq:amplitudes-OPE-TPE}) can be written as the following simple form.
\begin{eqnarray}
\mathcal{M}_{1\gamma} &\equiv& \mathcal{M}_{1\gamma}^{(a)} = c_1^{(1\gamma)} \mathcal{M}_1 + c_2^{(1\gamma)} \mathcal{M}_2,\nonumber \\
\mathcal{M}_{2\gamma,\pi} &\equiv& \mathcal{M}_{2\gamma,\pi}^{(a+b+c)} = c_{1\pi}^{(a+b+c)} \mathcal{M}_1 + c_{2\pi}^{(a+b+c)} \mathcal{M}_2,\nonumber\\
\mathcal{M}_{2\gamma,\rho} &\equiv& \mathcal{M}_{2\gamma,\rho}^{(a+b)} = c_{1\rho}^{(a+b)} \mathcal{M}_1 + c_{2\rho}^{(a+b)} \mathcal{M}_2,
\label{Eq:amplitude-definition}
\end{eqnarray}
with
\begin{eqnarray}
\mathcal{M}_1 &\equiv& i\bar{u}(p_3,m_e)(2\sla{p}_4+\sla{p}_3-\sla{p}_1)u(p_1,m_e) \  \bar{u}(p_5,m_n)\Gamma_5 u(p_2,m_p), \nonumber\\
\mathcal{M}_2 &\equiv& i\bar{u}(p_3,m_e)u(p_1,m_e) \  \bar{u}(p_5,m_n)\Gamma_5 u(p_2,m_p).
\label{Eq:invariant-amplitude}
\end{eqnarray}
The coefficients $c^{(1\gamma)}_{1,2}$ can be easily gotten which are expressed as
\begin{eqnarray}
c_1^{(1\gamma)} &=& \frac{4\pi  \alpha_e F_{\pi}(q^2)}{Q^2(t-m_{\pi}^2)}, \nonumber\\
c_2^{(1\gamma)} &=& 0,
\end{eqnarray}
with $\alpha_e\equiv e^2/4\pi$.

\section{Some analytic properties of the TPE contributions in $ep\rightarrow en\pi^+$}
\subsection{General properties due to the symmetry}
When taking the limit $m_e\rightarrow 0$, one has the following exact property due to the symmetry.
\begin{eqnarray}
c_{1\pi}^{(c)},c_{2\pi}^{(a+b)},c_{2\rho}^{(a+b)}\rightarrow 0.
\end{eqnarray}
Our manifest calculation also shows such property.

In the literature, the approximation $m_e=0$ is often used  before the loop integration since $m_e$ is much smaller than the other scales in the experimental region. In the elastic $ep$ scatting and elastic $e\pi$ scattering cases, one can find that such approach works well since the full TPE contributions are not dependent on $m_e$ at the leading order of $m_e$. In $ep\rightarrow en\pi^+$, we find that such approach is good for $c_{1\rho}^{(a+b)}$ but not good for $c_{1\pi}^{(a+b)}$. This is very different from the $ep$ or $e\pi$ cases and beyond the naive estimation. The detailed analytic calculation shows that there is a term like $\ln m_e$ in $c_{1\pi}^{(a+b)}$  when taking $m_e\rightarrow0$ after the loop calculation. Such term means that the usual Taylor series is not valid in the calculation. This is natural since the loop integration and the Taylor series is not commutated in some cases. Our numerical results also show such property and such log enhancement should be dealt carefully.

To keep this term, in the following calculation we at first take $m_e$ as non-zero and then expand the results on $m_e$. The packages FEYNCALC \cite{FenyCalc}, PackageX \cite{PackageX} and LOOPTOOL \cite{LoopTools}  are used in the practical calculations.

Under the pion-dominance approximation, although the cross sections are dependent on five variables but the TPE contributions $c_{1\pi,1\rho}^{(a,b)}$ are only dependent on three variables $t,Q^2$ and $\nu$. Due to the crossing symmetry, one has the following general relation when $Q^2$ and $t$ are fixed in the physical region:
\begin{eqnarray}
c_{1\pi,1\rho}^{(a)}(\nu^+,Q^2,t)&=&-c_{1\pi,1\rho}^{(b)}(-\nu^+,Q^2,t),
\label{Eq:relation-ab}
\end{eqnarray}
where $\nu^+=\nu+i0^+$.

\subsection{TPE contributions in the point-like particle case}
To show the analytic properties of the TPE contribution in a clear form, at first we take the point-like interaction as example. In this case, one has
\begin{eqnarray}
F^{\RM1}_{\gamma\pi\pi}(k^2)=F^{\RM1}_{\gamma\pi\rho}(k^2)=1,
\label{Eq:point-like}
\end{eqnarray}
where we have used the index $\RM1$ to refer to the point-like interaction. The same index is used for other quantities in the following expressions.

After the loop integration, we find the following analytic properties:

(1) There are no kinematic poles in $c_{1\pi,1\rho}^{\RM1,(a,b)}$.

\begin{figure}[htbp]
\centering
\includegraphics[height=6cm]{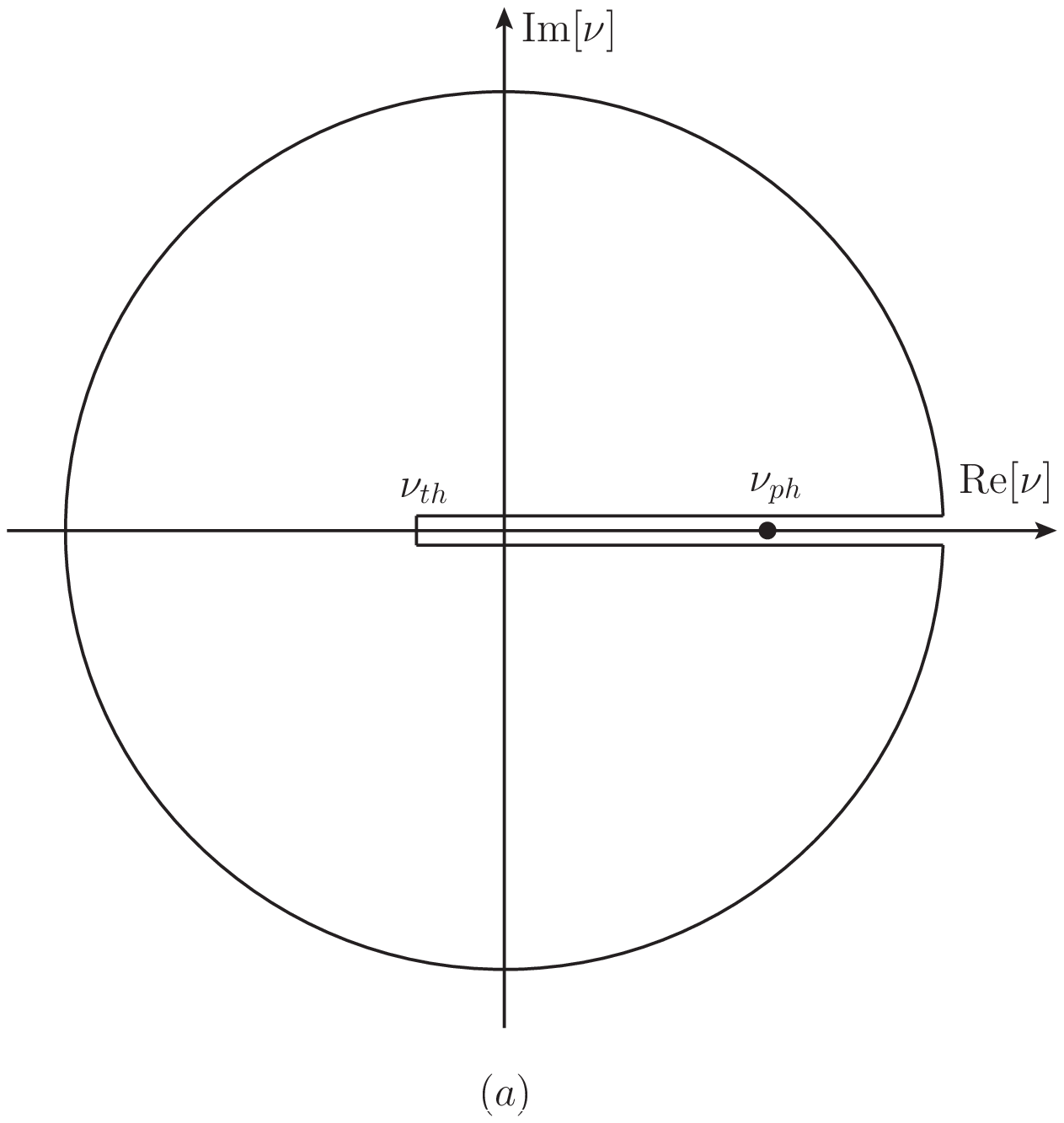}\includegraphics[height=6cm]{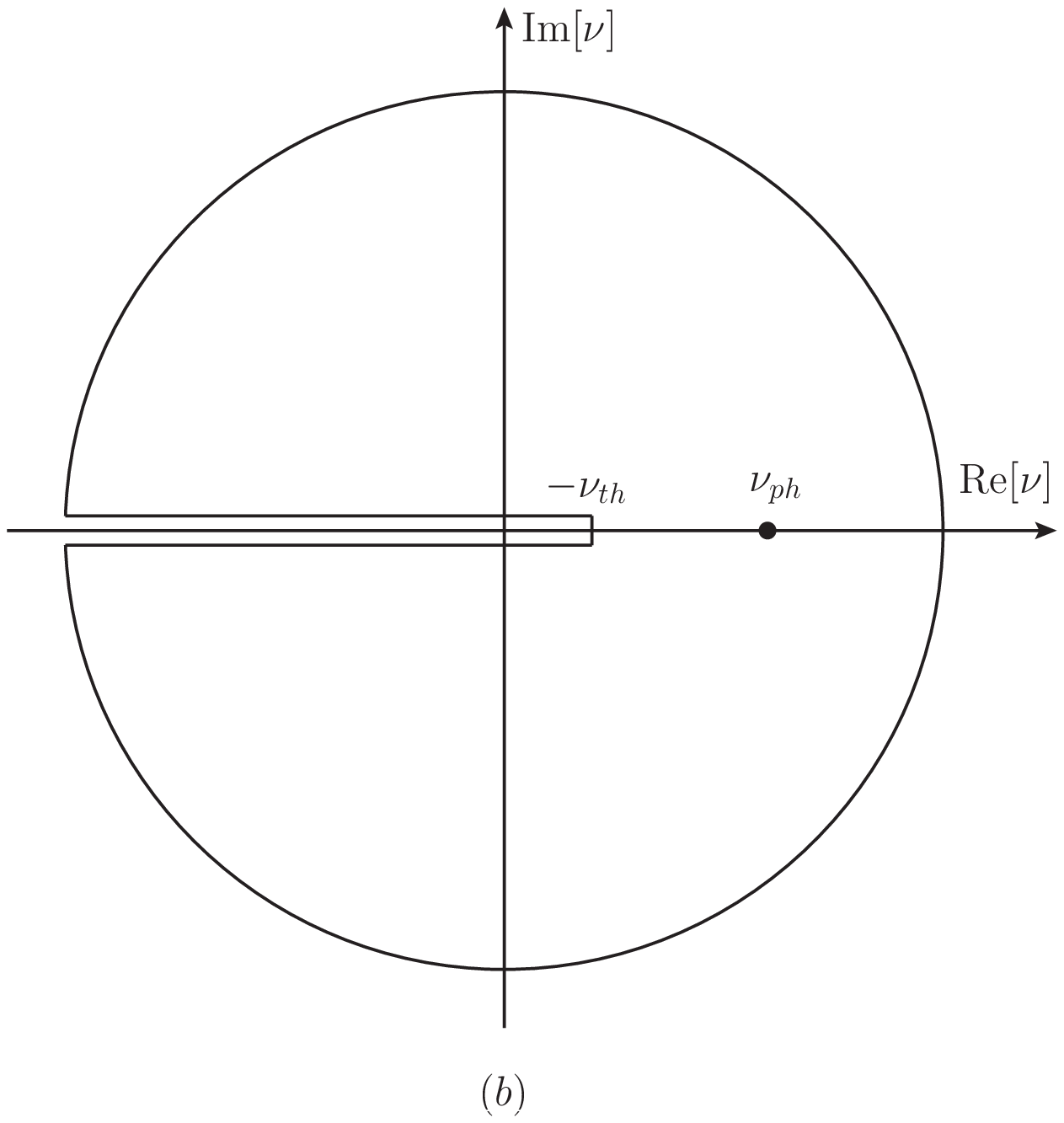}
\caption{The branch cuts of $c_{1\pi,1\rho}^{(a,b)}(\nu,Q^2,t)$ in the complex plane of $\nu$ at fixed physical $Q^2$ and $t$.  ($a$) is for $c_{1\pi,1\rho}^{\RM1,(a)}(\nu,Q^2,t)$ and ($b$) is for $c_{1\pi,1\rho}^{(b)}(\nu,Q^2,t)$ where $\nu_{th}^{(\pi)}=m_\pi^2+4m_em_\pi-Q^2-t$, $\nu_{th}^{(\rho)}=2m_\rho^2-m_\pi^2+4m_em_\rho-Q^2-t$ and $\nu_{ph}$ is the minimum physical $\nu$.
}
\label{Fig:Branch-cuts}
\end{figure}

(2) When  $t$ and $Q^2$  are fixed as physical values, the branch cuts of $c_{1\pi,1\rho}^{\RM1,(a,b)}$ on $\nu$ are showed as Fig. \ref{Fig:Branch-cuts}.

(3) The asymptotic behaviors of  $c_{1\pi,1\rho}^{\RM1,(a)}$ are expressed as follows:

\begin{eqnarray}
\textrm{Re}[c_{1\pi}^{\RM1,(a)}(\nu^+,Q^2,t] &\overset{\nu\rightarrow \infty}{\longrightarrow}& \frac{2\alpha_e^2}{Q^2(m_\pi^2-t)}\Big[\ln^2{\nu} -(\frac{1}{\widetilde{\epsilon}_{\textrm{IR}}}+\ln\frac{4m_e^2(m_\pi^2-t)^2\overline{\mu}^2_{\textrm{IR}}}{Q^4})\ln {\nu}+O(\nu^0)\Big],\nonumber\\
\textrm{Im}[c_{1\pi}^{\RM1,(a)}(\nu^+,Q^2,t] &\overset{\nu\rightarrow \infty}{\longrightarrow}& \frac{2\pi\alpha_e^2}{Q^2(m_\pi^2-t)}\Big[-2\ln\nu+(\frac{1}{\widetilde{\epsilon}_{\textrm{IR}}}
+\ln\frac{4m_e^2(m_\pi^2-t)^2\overline{\mu}^2_{\textrm{IR}}}{Q^4})+O(\nu^{-1})\Big], \nonumber\\
\label{Eq:asymptotic-pi-point-like}
\end{eqnarray}
and
\begin{eqnarray}
\textrm{Re}[c_{1\rho}^{\RM1,(a)}(\nu^+,Q^2,t] &\overset{\nu\rightarrow \infty}{\longrightarrow}& \frac{\alpha_e^2g_{\gamma\pi\rho}^2}{8m_\rho^2(m_\pi^2-t)}\Big[2\ln^2{\nu}-(4\ln Q^2+3+4\ln2)\ln{\nu}+O(\nu^0)\Big],\nonumber\\
\textrm{Im}[c_{1\rho}^{\RM1,(a)}(\nu^+,Q^2,t] &\overset{\nu\rightarrow \infty}{\longrightarrow}& \frac{\pi\alpha_e^2g_{\gamma\pi\rho}^2}{8m_\rho^2(m_\pi^2-t)}\Big[-4\ln\nu+(4\ln Q^2+3+4\ln2)+O(\nu^{-1})\Big],
\label{Eq:asymptotic-rho-point-like}
\end{eqnarray}
where $\overline{\mu}_{\textrm{IR}}$ is the $\textrm{IR}$ scale, and
\begin{eqnarray}
\frac{1}{\widetilde{\epsilon}_{\textrm{IR}}} &=& \frac{1}{\epsilon_{\textrm{IR}}}-\gamma_{E}+\ln4\pi. \nonumber
\end{eqnarray}
The asymptotical behaviors of $c_{1\pi,1\rho}^{\RM1,(b)}$ in the limit $\nu\rightarrow -\infty$ can be got easily via Eq. (\ref{Eq:relation-ab}).

Based on the above properties and Eqs. (\ref{Eq:relation-ab},\ref{Eq:asymptotic-pi-point-like},\ref{Eq:asymptotic-rho-point-like}), one can easily check that $c_{1\pi,1\rho}^{\RM1,(a)}$ satisfy the once-subtracted DR while $c_{1\pi,1\rho}^{\RM1,(a+b)}$ satisfy the following non-subtracted DR:
\begin{eqnarray}
\textrm{Re}[c_{1\pi,1\rho}^{\textrm{I},(a+b)}(\nu,Q^2,t)]
&=&\frac{2\nu}{\pi}\textrm{Re}\Big[\int_{\nu_{th}^{(\pi,\rho)}}^{\infty}\frac{\textrm{Im}[c_{1\pi,1\rho}^{\RM1,(a)}(\nu^+,Q^2,t)]}
{\overline{\nu}^2-\nu^2-i\epsilon}d\overline{\nu}\Big],
\label{Eq:DR}
\end{eqnarray}
where $\nu_{th}^{(\pi)}=m_\pi^2+4m_em_\pi-Q^2-t$, $\nu_{th}^{(\rho)}=2m_\rho^2-m_\pi^2+4m_em_\rho-Q^2-t$, and the manifest expressions for $\textrm{Im}[c_{1\pi,1\rho}^{\RM1,(a)}(\nu^+,Q^2,t)]$ are written as

\begin{eqnarray}
\textrm{Im}[c_{1\pi}^{\RM1,(a)}(\nu,Q^2,t)]{\Big |}_{m_e\rightarrow0}&=&\frac{2\pi\alpha_e^2}{Q^2(m_\pi^2-t)} \theta(x_1-4m_em_\pi) \nonumber\\  &&\times\Big[\frac{1}{\widetilde{\epsilon}_{\textrm{IR}}}+\ln\frac{4m_e^2(m_\pi^2-t)^2\overline{\mu}_{\textrm{IR}}^2}{x_1^2Q^4}
+\frac{2x_2}{x_2Q^2+y_1}\ln\frac{2x_2Q^2}{x_1x_3}\Big],
\label{Eq:expression-Im-pi-point-like}
\end{eqnarray}

and
\begin{eqnarray}
\textrm{Im}[c_{1\rho}^{\RM1,(a)}(\nu,Q^2,t)]{\Big |}_{m_e\rightarrow0} &=& \frac{\pi\alpha_e^2g_{\gamma\pi\rho}^2}{m_{\rho}^2(m_\pi^2-t)}\theta(x_4-4m_em_\pi)
\Big[\frac{x_4[3x_1x_2x_3-2m_{\rho}^2(2x_1x_2+2x_2x_3-x_1x_3)]}{8x_1x_2^2x_3}\nonumber\\
&&-\frac{x_2(\nu-2m_\rho^2)+2m_\pi^2t+2m_\rho^4}{2(x_2Q^2+y_1)}\ln\frac{2x_2Q^2}{x_1x_3}\Big],
\label{Eq:expression-Im-rho-point-like}
\end{eqnarray}

with
\begin{eqnarray}
x_1&=&Q^2+t+\nu-m_\pi^2, \nonumber\\
x_2&=&Q^2+t+\nu+m_\pi^2, \nonumber\\
x_3&=&Q^2-t+\nu+m_\pi^2, \nonumber\\
x_4&=&x_2-2m_\rho^2, \nonumber\\
y_1&=&m_\pi^4+m_\pi^2(Q^2-2t)+(t-\nu)(Q^2+t+\nu).
\end{eqnarray}
By the expressions of these imaginary parts and the DR, one can easily reproduce the real parts of $c_{1\pi,1\rho}^{\RM1,(a+b)}(\nu,Q^2,t)$.

We want to emphasize a general property that $\textrm{Im}[c_{1\pi}^{(a)}(\nu,Q^2,t)]$  has IR divergence and is dependent on the IR scale $\overline{\mu}_{\textrm{IR}}$. This is natural since the DR Eq.(\ref{Eq:DR}) means that $\textrm{Re}[c_{1\pi}^{(a+b)}(\nu,Q^2,t)]$ is totally determined by $\textrm{Im}[c_{1\pi}^{(a)}(\nu,Q^2,t)]$ and the former has IR divergence. This property hints that $\textrm{Im}[c_{1\pi}^{(a)}(\nu,Q^2,t)]$ can not be determined by experimental data directly. This is very different with the case in the forward angle limit. On the contrary, the contribution $\textrm{Im}[c_{1\rho}^{(a)}(\nu,Q^2,t)]$ has no IR divergence.

\subsection{TPE contributions with EM FFs}
Physically, the EM FFs $F_{\gamma\pi\pi}$ and $F_{\gamma\pi\rho}$ are not constants and the momentum dependence of the EM FFs should be considered when $Q^2$ increases. In the practical calculation, for simplicity the following monopole form FF is used \cite{Blunden2010-pion-form-factor,zhouhq2011-pion-photon-interaction}.
\begin{eqnarray}
F_{\gamma\pi\pi}^{\RM2}(q^2)=F_{\gamma\pi\rho}^{\RM2}(q^2)=\frac{-\Lambda^2}{q^2-\Lambda^2}.
\end{eqnarray}

After the loop integration with this FF as inputs, we find the properties on the kinematic poles and the branch cuts of $c_{1\pi,1\rho}^{\RM2,(a,b)}$ are the same with those of $c_{1\pi,1\rho}^{\RM1,(a,b)}$. The asymptotic behaviors of $c_{1\pi,1\rho}^{\RM2,(a)}$ are expressed as follows:
\begin{eqnarray}
\textrm{Re}[c_{1\pi}^{\RM2,(a)}(\nu^+,Q^2,t)] &\overset{\nu\rightarrow \infty}{\longrightarrow}& \frac{2\alpha_e^2}{Q^2(m_\pi^2-t)}\Big[\frac{\Lambda^2}{\Lambda^2+Q^2}\ln^2\nu+a_{1\pi}^{\RM2}\ln{\nu}+O(\nu^{0})\Big],\nonumber\\
\textrm{Im}[c_{1\pi}^{\RM2,(a)}(\nu^+,Q^2,t)] &\overset{\nu\rightarrow \infty}{\longrightarrow}&
\frac{2\pi\alpha_e^2}{Q^2(m_\pi^2-t)}\Big[\frac{-2\Lambda^2}{\Lambda^2+Q^2}\ln{\nu}-a_{1\pi}^{\RM2}+O(\nu^{-1})\Big],
\label{Eq:asymptotic-pi-with-FF}
\end{eqnarray}
and
\begin{eqnarray}
\textrm{Re}[c_{1\rho}^{\RM2,(a)}(\nu^+,Q^2,t)] &\overset{\nu\rightarrow \infty}{\longrightarrow}&a_{1\rho}^{\RM2}\ln{\nu}+O(\nu^0), \nonumber\\
\textrm{Im}[c_{1\rho}^{\RM2,(a)}(\nu^+,Q^2,t)] &\overset{\nu\rightarrow \infty}{\longrightarrow}&-2\pi a_{1\rho}^{\RM2}+O({\nu}^{-1}),
\label{Eq:asymptotic-rho-with-FF}
\end{eqnarray}
where $a_{1\pi,1\rho}^{\RM2}$ are functions only dependent on $m_\pi,m_\rho,t,Q^2$ and $\Lambda$.

Comparing the asymptotic behaviors Eqs. (\ref{Eq:asymptotic-pi-with-FF},\ref{Eq:asymptotic-rho-with-FF}) and Eqs. (\ref{Eq:asymptotic-pi-point-like},\ref{Eq:asymptotic-rho-point-like}), one can find an interesting property: the asymptotic behaviors of $c_{1\pi}^{\RM2,(a)}(\nu^+,Q^2,t)$ is similar with $c_{1\pi}^{\RM1,(a)}(\nu^+,Q^2,t)$, but the asymptotic behaviors of $c_{1\rho}^{\RM2,(a)}(\nu^+,Q^2,t)$ is a little different from $c_{1\rho}^{\RM1,(a)}(\nu^+,Q^2,t)$. The asymptotic behavior of $\textrm{Im}[c_{1\pi}^{\RM1,(a)}(\nu^+,Q^2,t)]/\textrm{Im}[c_{1\rho}^{\RM1,(a)}(\nu^+,Q^2,t)]$ is $\sim 1$  while $\textrm{Im}[c_{1\pi}^{\RM2,(a)}(\nu^+,Q^2,t)]/\textrm{Im}[c_{1\rho}^{\RM2,(a)}(\nu^+,Q^2,t)]$ is $\sim\ln\nu$. This property directly means that when $\nu$ increases the contributions with one $\rho$ intermediate state is suppressed by a factor $\ln\nu$  in the monopole FF case, while there is no such factor in the point like case.

The above properties also show that the TPE contributions $c_{1\pi,1\rho}^{\RM2,(a+b)}(\nu,Q^2,t)$ still satisfy the DR Eq. (\ref{Eq:DR}) when $m_e\rightarrow 0$.  Since the analytic expressions can be used directly and conveniently to analysis the further experimental data.  Here we list the expressions for the imaginary parts when $m_{e}\rightarrow 0$. After separating the TPE contribution from  $\pi$ intermediate into two parts as
\begin{eqnarray}
c_{1\pi}^{\RM2,(a)}(\nu,Q^2,t)&\equiv& c_{1\pi,\textrm{IR}}^{\RM2,(a)}(\nu,Q^2,t)+  c_{1\pi,\textrm{fin}}^{\RM2,(a)}(\nu,Q^2,t),
\end{eqnarray}
the imaginary parts of the TPE contributions from one $\pi$, one $\rho$ intermediate state are expressed as follows:
\begin{eqnarray}
\textrm{Im}[c_{1\pi,\textrm{IR}}^{\RM2,(a)}(\nu,Q^2,t)]{\Big |}_{m_e\rightarrow0}&=&\frac{2\pi\alpha_e^2}{Q^2(m_\pi^2-t)} \theta(x_1-4m_em_\pi) \frac{\Lambda^2}{\Lambda^2+Q^2}\frac{1}{{\widetilde{\epsilon}_{\textrm{IR}}}}, \nonumber\\
\textrm{Im}[c_{1\pi,\textrm{fin}}^{\RM2,(a)}(\nu,Q^2,t)]{\Big |}_{m_e\rightarrow0}&=&\frac{2\pi\alpha_e^2}{Q^2(m_\pi^2-t)} \theta(x_1-4m_em_\pi) \sum_{i=1}^{5}g_{\pi,i}\log z_{\pi,i},\nonumber\\
\textrm{Im}[c_{1\rho}^{\RM2,(a)}(\nu,Q^2,t)]{\Big |}_{m_e\rightarrow0}&=&\frac{2\pi\alpha_e^2g_{\gamma\pi\rho}^2}{m_{\rho}^2(m_{\pi}^2-t)(y_1+Q^2x_2)} \theta(x_4-4m_em_\pi) \sum_{i=1}^{5}g_{\rho,i}\log z_{\rho,i},
\label{Eq:Im-part-c1-with-FF}
\end{eqnarray}
where
\begin{eqnarray}
g_{\pi,1} &=& \frac{\Lambda^2}{\Lambda^2+Q^2}, \nonumber\\
g_{\pi,2} &=&\frac{\Lambda^2x_3}{Q^2x_1+\Lambda^2x_3},\nonumber\\
g_{\pi,3}&=&  \frac{Q^2(\Lambda^2x_2-y_1)}{(\Lambda^2+Q^2)(Q^2x_2+y_1) },\nonumber\\
g_{\pi,4}&=&  \frac{Q^2(x_1y_1-\Lambda^2x_2x_3)}{(Q^2x_2+y_1)(Q^2x_1+\Lambda^2x_3)},\nonumber\\
g_{\pi,5}&=&  \frac{Q^2[2\Lambda^2x_2(Q^2+\nu)-x_1y_1]}{(Q^2x_2+y_1)y_2},
\end{eqnarray}
\begin{eqnarray}
z_{\pi,1}&=& \frac{2\overline{\mu}^2_{\textrm{IR}}x_2}{x_1^2}, \nonumber\\
z_{\pi,2}&=& \frac{2 m_e^2(m_\pi^2-t)^2}{Q^4x_2}, \nonumber\\
z_{\pi,3}&=& \frac{\Lambda^2(2\Lambda^2x_2+x_1x_3)}{2x_2(\Lambda^2+Q^2)^2}, \nonumber\\
z_{\pi,4}&=& \frac{x_1^2(Q^2x_1+\Lambda^2x_3)^2}{2\Lambda^2Q^4x_2(x_1^2+2\Lambda^2x_2)}, \nonumber\\
z_{\pi,5}&=& \frac{h_1+x_1y_2}{h_1-x_1y_2},
\end{eqnarray}
\begin{eqnarray}
g_{\rho,1} &=&-\frac{1}{4}\Lambda^2m_{\rho}^2,\nonumber\\
g_{\rho,2} &=& \frac{1}{8}(y_3-m_\pi^2x_5), \nonumber\\
g_{\rho,3}&=&  -\frac{1}{8}(y_3-m_{\pi}^2x_5+2\Lambda^2m_\rho^2),\nonumber\\
g_{\rho,4}&=&  \frac{1}{4}\Lambda^2(m_{\rho}^2-\Lambda^2),\nonumber\\
g_{\rho,5}&=&  \frac{1}{8y_4}\Big[Q^2x_4(y_3-m_\pi^2x_5)+\Lambda^2(h_2+m_\pi^2y_5)+2\Lambda^4y_6-4\Lambda^6(Q^2+\nu)\Big],
\end{eqnarray}
and
\begin{eqnarray}
z_{\rho,1}&=& \frac{64x_2^2(m_\rho^2-m_\pi^2)^2(m_\rho^2-t)^2}{x_1^2x_3^2x_4^2}, \nonumber\\
z_{\rho,2}&=& \frac{Q^4x_4^2}{16(m_\rho^2-m_\pi^2)^2(m_\rho^2-t)^2}, \nonumber\\
z_{\rho,3}&=& \frac{x_4^2(Q^2x_4+\Lambda^2x_1)^2(Q^2x_4+\Lambda^2x_3)^2}{16\Lambda^4(m_\rho^2-m_\pi^2)^2(m_\rho^2-t)^2
\big(x_1x_3x_4^2+4\Lambda^2x_2[x_4(Q^2+\nu)+\Lambda^2x_2]\big)}, \nonumber\\
z_{\rho,4}&=& \frac{4\Lambda^4x_2^2}{x_1x_3x_4^2+4\Lambda^2x_2[x_4(Q^2+\nu)+\Lambda^2x_2]}, \nonumber\\
z_{\rho,5}&=& \frac{x_4(Q^2x_4+y_4)+2\Lambda^2x_4(Q^2+\nu)+4\Lambda^4x_2}{x_4(Q^2x_4-y_4)+2\Lambda^2x_4(Q^2+\nu)+4\Lambda^44x_2},
\end{eqnarray}
with
\begin{eqnarray}
x_5&=&\nu+t-2m_\rho^2,\nonumber\\
y_2&=&\sqrt{4 \Lambda ^4[m_\pi^4+2m_\pi^2(Q^2-t)+2Q^4+2Q^2(t+\nu)+t^2]+4\Lambda ^2 Q^2(Q^2+\nu) x_1+Q^4 x_1^2},\nonumber\\
y_3&=&-2m_\rho^4+(2m_\rho^2-\nu)(Q^2+t+\nu),\nonumber\\
y_4&=&\sqrt{4\Lambda^4[m_\pi^4+2m_\pi^2(Q^2-t)+2Q^4+2Q^2(t+\nu)+t^2]+4\Lambda^2Q^2(Q^2+\nu)x_4+Q^4x_4^2},\nonumber\\
y_5&=&4m_\rho^2(2Q^2+\nu)-2(Q^2+\nu)(2t+\nu),\nonumber\\
y_6&=&2m_\pi^4+m_\pi^2(3Q^2-4t)+2m_\rho^2(3Q^2+2\nu)+(Q^2+2t-2\nu)(Q^2+t+\nu),\nonumber\\
h_1&=&4\Lambda^4x_2+2\Lambda^2(Q^2+\nu)x_1+Q^2x_1^2,\nonumber\\
h_2&=&-4m_\rho^4(3Q^2+\nu)+4m_\rho^2(2Q^2+\nu)(Q^2+t+\nu)-2\nu(Q^2+\nu)(Q^2+t+\nu ).
\end{eqnarray}

We also want to  point out that the contributions $c_{2\pi,2\rho}^{\RM2,(a+b)}$ are non-zero and do not satisfy the DR  Eq. (\ref{Eq:DR}) when $m_e$ is taken as non-zero.

Finally, the real parts of the TPE contributions $\textrm{Re}[c_{1\pi,1\rho}^{\RM2,(a+b)}(\nu,Q^2,t)]$ can be got directly and easily from Eq.(\ref{Eq:DR}) and Eq.(\ref{Eq:Im-part-c1-with-FF}).

\section{The numerical results for $c_{1\pi,1\rho}^{\RM2,(a+b)}/c_1^{(1\gamma)}$}
Usually, the experimental quantities $Q^2,W,\epsilon,\theta_\pi$ and $\phi_\pi$ are chosen as variables to express the differential cross section where $\epsilon$ is the virtual photon polarization, $\theta_\pi$ and $\phi_\pi$ are the angles between the three-momentum of $\pi$ and the $ep$ scattering plane. Their detailed definitions can be found in the Appendix of Ref. \cite{zhouhq2020-TPE-pi-intermediate}.
In the poin-exchange dominance approximation, as discussed above the coefficients of the invariant amplitudes are only dependent on $\nu,t$ and $Q^2$ when taking $Q^2, W, \nu, t$ and $s$ as five independent variables. This property means it is much simpler to show the TPE contributions by choosing the latter quantities as independent variables. In the following, at first we present the numeric results with the latter choice, and then present the numeric results with the experimental variables as inputs.

\begin{figure}[htbp]
\centering
\includegraphics[width=15cm]{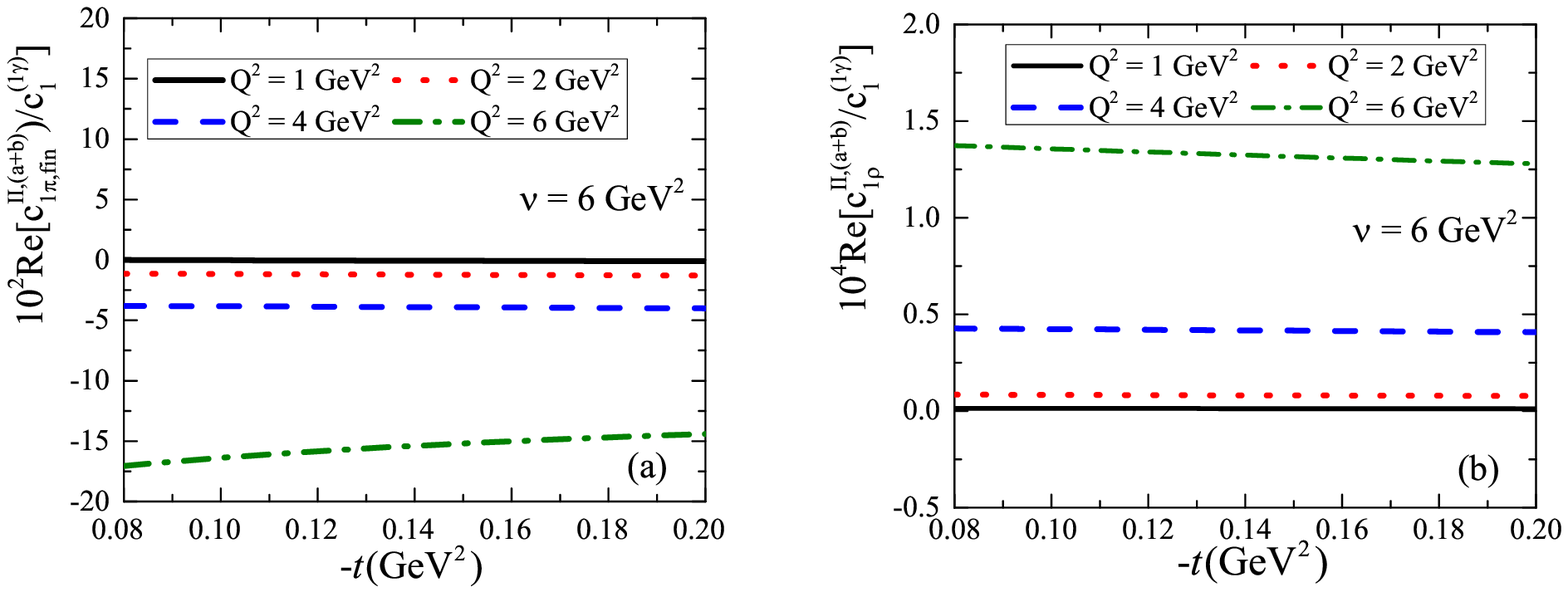}
\includegraphics[width=15cm]{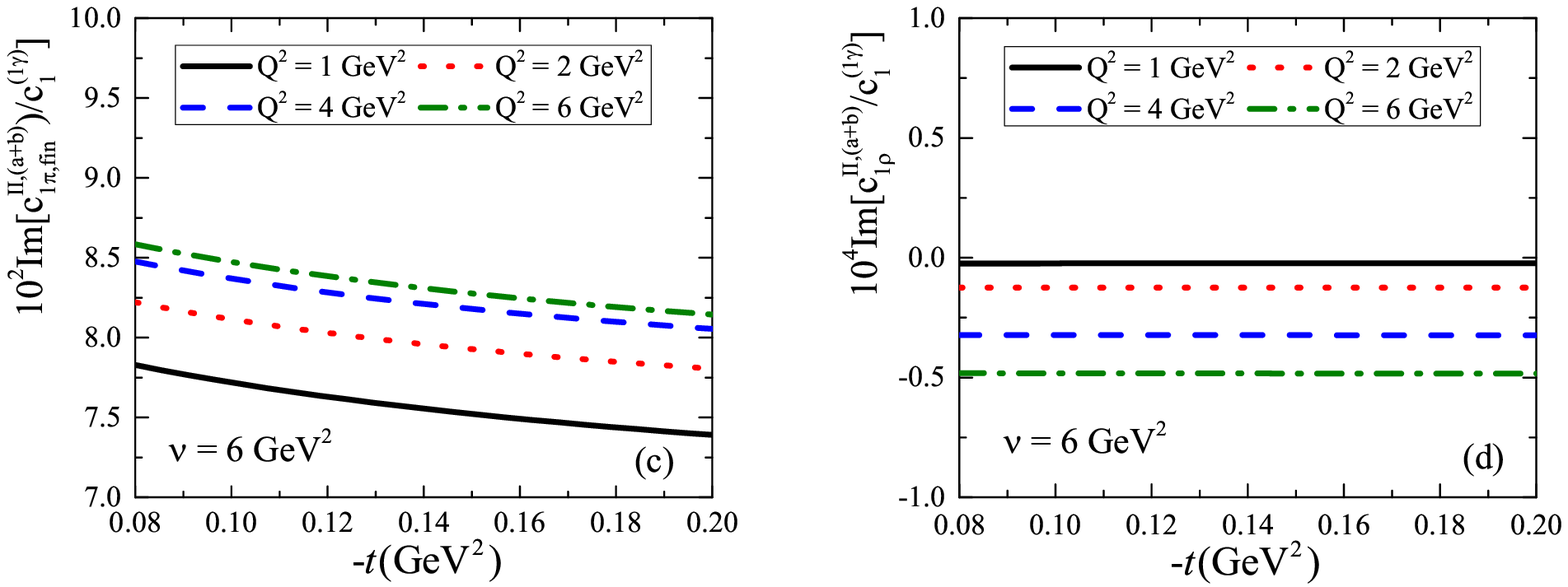}
\caption{Numeric results for $c_{1\pi,\textrm{fin}}^{\RM2,(a+b)}/c_1^{(1\gamma)}$ \textit{vs.} $-t$ and $c_{1\rho}^{\RM2,(a+b)}/c_1^{(1\gamma)}$  \textit{vs.} $-t$ at fixed $Q^2, \nu$. The top panel is for the real part and the bottom panel is for the imaginary part.}
\label{Fig:ratio-t-fixed-Qv}
\end{figure}

In Fig. \ref{Fig:ratio-t-fixed-Qv}, the numeric results for $c_{1\pi,\textrm{fin}}^{\RM2,(a+b)}/c_1^{(1\gamma)}$ and $c_{1\rho}^{\RM2,(a+b)}/c_1^{(1\gamma)}$ are presented at $Q^2=1,2,4,6$ GeV$^2$ with $\nu=6$ GeV$^2$. Here we have taken the IR scale as $\overline{\mu}_{\textrm{IR}}=1$ GeV, the parameter in the FFs as $\Lambda=0.77$ GeV and the coupling constant as $g_{\gamma\pi\rho}=0.103$ which is determined by the decay width of $\rho^+\rightarrow \gamma\pi^+$. The numeric results clearly show that the TPE contributions with one $\rho$ meson intermediate state are much smaller than those with one $\pi$ intermediate state in the chosen regions. This interesting property is very different from the property in elastic $ep$ scattering case where the contributions with inelastic intermediate states are at the same order with those with an elastic intermediate state. The numeric results also show that the absolute magnitudes of the TPE corrections increase when $Q^2$ increases. When $\nu=6$ GeV$^2$, the magnitudes of  $\textrm{Re}[c_{1\pi,\textrm{fin}}^{\RM2,(a+b)}/c_1^{(1\gamma)}]$ and $\textrm{Re}[c_{1\rho}^{\RM2,(a+b)}/c_1^{(1\gamma)}]$ are about $15\%$ and $0.15\%$ at $Q^2=6$ GeV$^2$, about $4\%$ and $0.005\%$ at $Q^2=4$ GeV$^2$, respectively. Another very interesting property is that the TPE corrections are not sensitive on the variable $t$ when $\nu$ and $Q^2$ are fixed.

\begin{figure}[htbp]
\centering
\includegraphics[width=15cm]{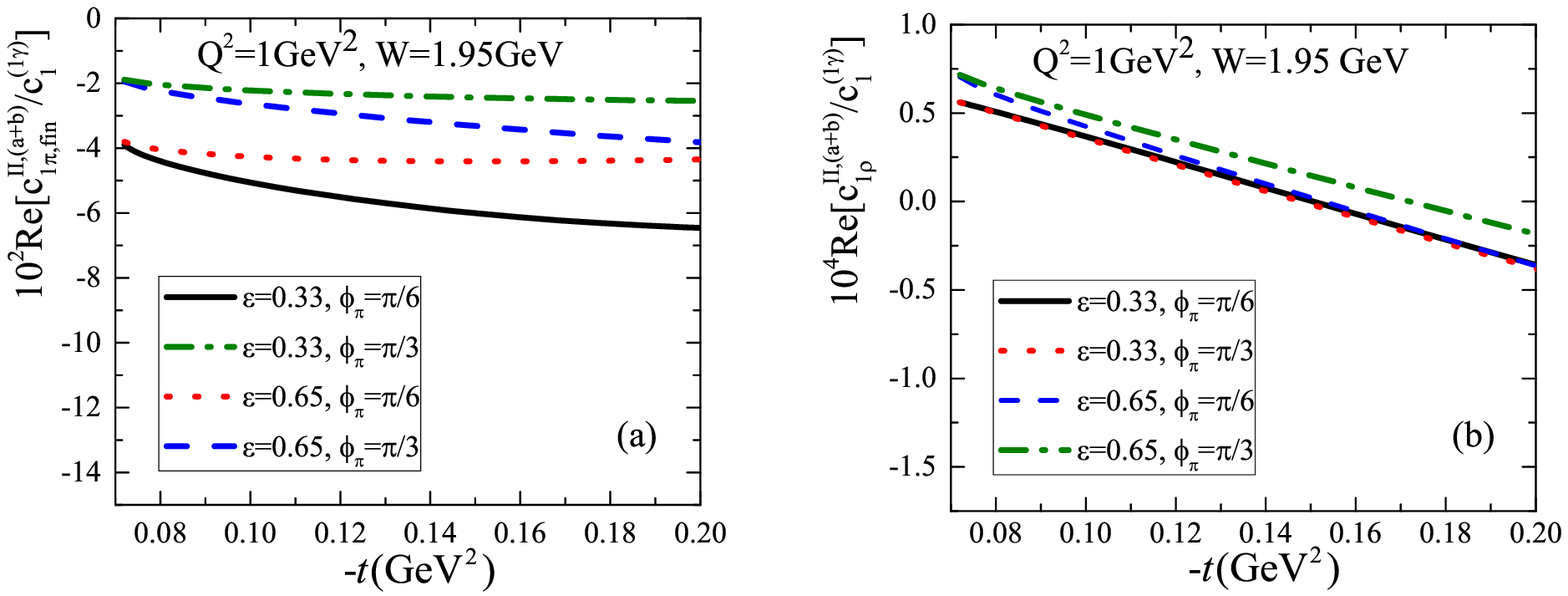}
\includegraphics[width=15cm]{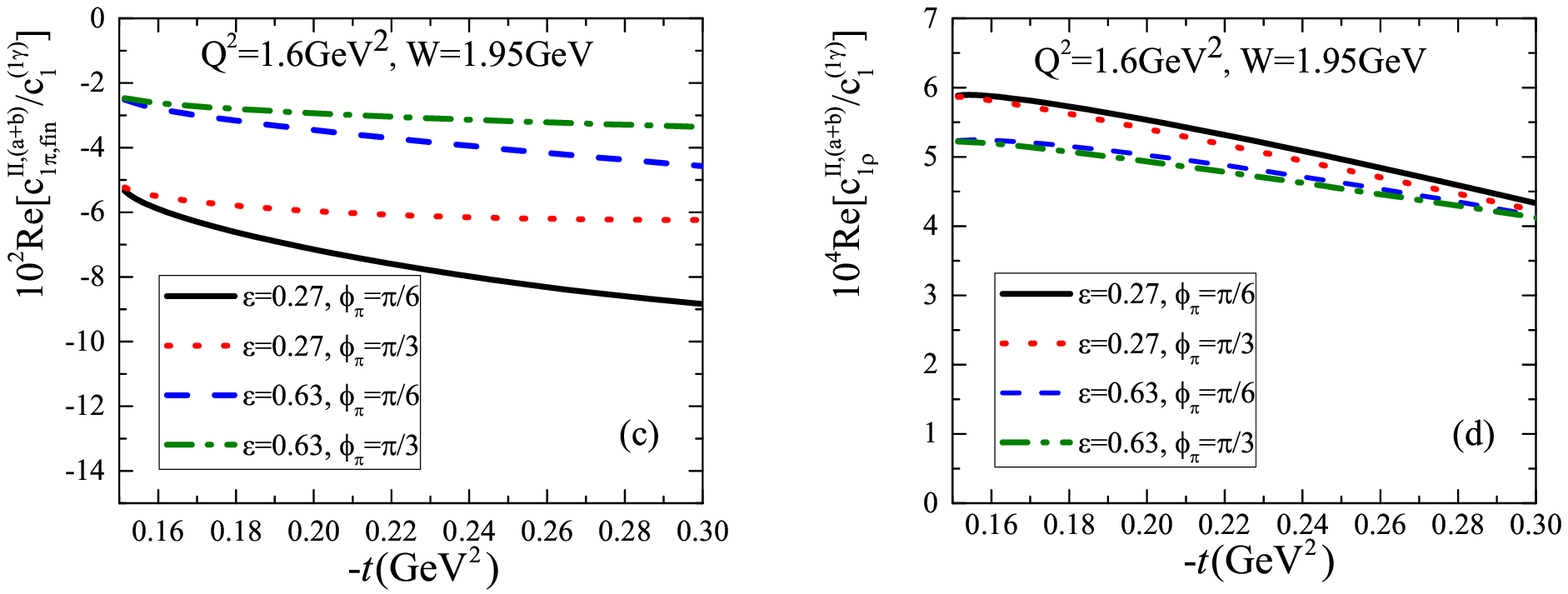}
\caption{Numeric results for $\textrm{Re}[c_{1}^{((\rho,a+b))}/c_1^{(1\gamma)}]$ \textit{vs.} $-t$ at fixed $Q^2,W,\epsilon$ and $\phi_\pi$. This is the result with $Q^2=1.6$ GeV$^2$.}
\label{Fig:Re-ratio-t-rho-Ex}
\end{figure}

In Fig. \ref{Fig:Re-ratio-t-rho-Ex}, we take the kinematics in JLab $F_\pi$ experiment \cite{electro-pion-production-JLab-1} with $Q^2=1,1.6 \text{ GeV}^2$ at $W=1.95 \text{ GeV}$ as examples to show the numerical results for $\textrm{Re}[c_{1\pi,\textrm{fin}}^{\RM2,(a+b)}/c_1^{(1\gamma)}]$ and $\textrm{Re}[c_{1\rho}^{\RM2,(a+b)}/c_1^{(1\gamma)}]$. The numerical results for $\textrm{Re}[c_{1\pi,\textrm{fin}}^{\RM2,(a+b)}/c_1^{(1\gamma)}]$ are the same as those in Ref. \cite{zhouhq2020-TPE-pi-intermediate} where they are labelled as $\textrm{Re}[c_{1\pi,\textrm{fin}}^{(2\gamma)}/c_1^{(1\gamma)}]$. The (blue) dashed curves and the (olive) dash-dotted  curves refer to the results at $\phi_{\pi}=\pi/6$  and $\phi_{\pi}=\pi/3$ with $\epsilon=0.65$ or $0.63$, the (black) solid curves and the (red) dotted curves are associated with $\epsilon=0.33$ or $0.27$.  The results clearly show that the absolute magnitude of TPE corrections $\textrm{Re}[c_{1\rho}^{\RM2,(a+b)}/c_1^{(1\gamma)}]$ are smaller than $10^{-4}$ at $Q^2=1$ GeV$^2$ and smaller than $10^{-3}$ at $Q^2=1.6$ GeV$^2$. These corrections are much smaller than the results with one $\pi$ meson intermediate state \cite{zhouhq2020-TPE-pi-intermediate}. In the practical estimation, one can use  Eq.(\ref{Eq:DR}) and Eq.(\ref{Eq:Im-part-c1-with-FF}) to get $\textrm{Re}[c_{1\pi,1\rho}^{\RM2,(a+b)}]$ easily and then check this property in a large region. This property suggests that the TPE corrections with one $\rho$ meson intermediate state can be relatively negligible in the current experimental regions.


In summary, in this work the TPE contributions in $ep\rightarrow en\pi^+$ with one $\pi$, one $\rho$ intermediate state are estimated under the
pion-dominance approximation within the hadronic model. The calculation shows that these TPE contributions satisfy an un-subtracted DR when $m_e\rightarrow 0$. The analytic expressions for the imaginary parts of the TPE contributions within the hadronic model are given. Combine these analytic expressions and the DR, one can get the real parts of the TPE contributions at any available kinematic region easily. We think these expression can help the further experimental analysis to include the TPE contributions conveniently. On the numerical part, we find the contributions from one $\rho$ intermediate state are much smaller than those from  one $\pi$ intermediate state. This suggests that the estimation only with one $\pi$ intermediate state can be applied to higher $Q^2$ and higher $\nu$ safely.

\section{Acknowledgments}

The author Hai-Qing Zhou would like to thank Hiren. Pate for his kind help in PackageX.  This work is supported by the  National Natural Science Foundations of China under Grant No. 12075058 and No. 11975075.


\begin{thebibliography}{99}

\bibitem{hadronic model}
P. G. Blunden, W. Melnitchuk, and J. A. Tjon, Phys. Rev.Lett. {\bf 91}, 142304 (2003);
S. Kondratyuk, P. G. Blunden, W. Melnitchuk, and J. A.Tjon, Phys. Rev. Lett. {\bf 95}, 172503 (2005);
P. G. Blunden, W. Melnitchuk, and J. A. Tjon, Phys. Rev. C {\bf 72}, 034612 (2005).

\bibitem{GPD method}
Y. C. Chen, A. Afanasev, S. J. Brodsky, C. E. Carlson, and M. Vanderhaeghen, Phys. Rev. Lett. {\bf 93}, 122301 (2004);
A. Afanasev, S. J. Brodsky, C. E. Carlson, Y. C. Chen, and M. Vanderhaeghen, Phys. Rev. D {\bf 72}, 013008 (2005).

\bibitem{pQCD method}
N. Kivel and M. Vanderhaeghen, Phys. Rev. Lett. {\bf 103}, 092004 (2009);
D. Borisyuk and A. Kobushkin, Phys. Rev. C {\bf 79}, 034001 (2009).



\bibitem{dispersion relation-1}
D. Borisyuk and A. Kobushkin, Phys. Rev. C {\bf 74}, 065203 (2006);
D. Borisyuk and A. Kobushkin, Phys. Rev. C {\bf 83}, 025203 (2011);
D. Borisyuk and A. Kobushkin, Phys. Rev. C {\bf 86}, 055204 (2012);
D. Borisyuk and A. Kobushkin, Phys. Rev. C {\bf 89}, 025204 (2014);

\bibitem{dispersion relation-2}
O. Tomalak and M. Vanderhaeghen, Eur. Phys. J. A {\bf 51}, 24 (2015);
P. G. Blunden and W. Melnitchouk, Phys. Rev. C {\bf 95}, 065209 (2017).
Oleksandr Tomalak, Barbara Pasquini, and Marc Vanderhaeghen, Phys. Rev. D {\bf96},096001 (2017)


\bibitem{SCEF}
N. Kivel and M. Vanderhaeghen, J. High Energy Phys. {\bf 04}, 029 (2013).

\bibitem{TPE-ep-ChpT-2020-2021}
P.~Talukdar, V.~C.~Shastry, U.~Raha and F.~Myhrer,
Phys. Rev. D \textbf{101}, no.1, 013008 (2020);

P.~Talukdar, V.~C.~Shastry, U.~Raha and F.~Myhrer,
Phys. Rev. D \textbf{104}, no.5, 053001 (2021)
doi:10.1103/PhysRevD.104.053001
[arXiv:2010.09380 [nucl-th]].

Xiong-Hui Cao, Qu-Zhi Li and Han-Qing Zheng, arXiv:2112.06230 [hep-ph].





\bibitem{phenomenological parametrizations}
Y. C. Chen, C. W. Kao, and S. N. Yang, Phys. Lett. B {\bf 652}, 269 (2007);
D. Borisyuk and A. Kobushkin, Phys. Rev. C {\bf 76}, 022201 (2007).



\bibitem{VEPP3-2012}
A. V. Gramolin {\it et al.}, Nucl. Phys. Proc. Suppl. \textbf{ 225-227}, 216 (2012).

\bibitem{VEPP3-2014}
D. M. Nikolenko {\it et al.}, EPJ Web of Conferences \textbf{ 66}, 06002 (2014).


\bibitem{Moteabbed13}
M. Moteabbed {\it et al.} (CLAS Coll.),  Phys. Rev. C \textbf{ 88}, 025210 (2013),


\bibitem{Kohl14}
M. Kohl {\it et al.} (OLYMPUS Coll.),  EPJ Web Conf. \textbf{ 66}, 06009  (2014).


\bibitem{electro-pion-production-Cornell}
C. J. Bebek \textit{et al}., Phys. Rev. D {\bf 13}, 25 (1976);
C. J. Bebek \textit{et al}., Phys. Rev. D {\bf 17}, 1693 (1978).

\bibitem{electro-pion-production-DESY}
P. Brauel \textit{et al}. (DESY), Phys. Lett. B {\bf 65}, 184 (1976);
P. Brauel \textit{et al}., Phys. Lett. B {\bf 69}, 253 (1977);
H. Ackermann \textit{et a}l., Nucl. Phys. B {\bf 137}, 294 (1978);
P. Brauel \textit{et al}., Z. Phys. C {\bf 3}, 101 (1979).


\bibitem{electro-pion-production-JLab-1}
J. Volmer \textit{et al}. (Jefferson Lab F$\pi$ Collaboration), Phys. Rev. Lett. {\bf 86}, 1713 (2001);
T. Horn \textit{et a}l. (Jefferson Lab F$\pi$ Collaboration), Phys. Rev. lett. {\bf 97}, 192001 (2006);
V. Tadevosyan \textit{et al}. (Jefferson Lab F$\pi$ Collaboration), Phys. Rev. C {\bf 75}, 055205 (2007).

\bibitem{electro-pion-production-JLab-2}
H. P. Blok, T. Horn \textit{et al}. (Jefferson Lab F$\pi$ Collaboration), Phys. Rev. C {\bf 78}, 045202 (2008).


\bibitem{electro-pion-production-JLab-3}
G.M. Huber,\textit{ et al}. (Jefferson Lab F$\pi$ Collaboration), Phys. Rev. C {\bf 78}, 045203 (2008).





\bibitem{Afanasev2013}
Andrei Afanasev, Aleksandrs Aleksejevs and Svetlana Barkanova, Phys. Rev. C {\bf 88},053008 (2013).


\bibitem{zhouhq2020-TPE-pi-intermediate}
Hui Yun Cao, Hai Qing Zhou, Phys. Rev. C {\bf 101}, 055201 (2020).





\bibitem{zhouhq2011-pion-photon-interaction}
Hai Qing Zhou, Phys. Lett. B {\bf 706}, 82-85, (2011).

\bibitem{g-rho-pi-gamma}
T. Feuster and U. Mosel, Phys. Rev. C {\bf 59}, 460-491, (1999).





\bibitem{FenyCalc}
R. Mertig, M. Bohm and Ansgar Denner, Comput. Phys. Commun. {\bf 64}, 345 (1991);
Vladyslav Shtabovenko, Rolf Mertig and Frederik Orellana, Comput. Phys. Commun. {\bf 207}, 432 (2016).

\bibitem{PackageX}
H. H. Patel,  Comput. Phys. Commun. {\bf 197}, 276-290 (2015);
H. H. Patel,  Comput. Phys. Commun. {\bf 218}, 66-70 (2017).

\bibitem{LoopTools}
T. Hahn, M. Perez-Victoria, Comput. Phys. Commun. {\bf 118}, 153 (1999).


\bibitem{Blunden2010-pion-form-factor}
P. G. Blunden, W. Melnitchouk, J. A. Tjon, Phys. Rev. C {\bf 81}, 018202 (2010).













\end{thebibliography}
\end{document}